\documentclass[aps,pre,twocolumn]{revtex4-2}
\usepackage{amssymb}
\usepackage{amsmath,bm}
\usepackage{graphicx,color}
\usepackage{bbm}
\usepackage{multirow}
\usepackage{ragged2e}
\usepackage{hyperref}
\hypersetup{colorlinks=true,citecolor=blue,linkcolor=red,urlcolor=blue}
\usepackage{color}

\definecolor{green}{rgb}{0,0.5,0}

\newcommand{\B}[1]{{\bm{#1}}}

\begin{document}

\title{Universal Density of Low Frequency States in Silica Glass at Finite Temperatures}
\author{Roberto Guerra$^1$, Silvia Bonfanti$^1$, Itamar Procaccia$^{2,3}$ and Stefano Zapperi$^{1,4}$ }

\affiliation{
	$^1$Center for Complexity and Biosystems, Department of Physics, University of Milan, via Celoria 16, 20133 Milano, Italy
	\\$^2$ Dept. of Chemical Physics, The Weizmann Institute of Science, Rehovot 76100, Israel
	\\$^3$Center for OPTical IMagery Analysis and Learning, Northwestern Polytechnical University, Xi'an, 710072 China
	\\$^4$ CNR - Consiglio Nazionale delle Ricerche,  Istituto di Chimica della Materia Condensata e di Tecnologie per l'Energia, Via R. Cozzi 53, 20125 Milano, Italy
}

\date{\today}

\begin{abstract}
The theoretical understanding of the low-frequency modes in amorphous solids at finite temperature is still incomplete. The study of the relevant modes is obscured by the dressing of inter-particle forces by collision-induced momentum transfer that is unavoidable at finite temperatures. Recently, it was proposed that low frequency modes of vibrations around the {\em thermally averaged} configurations deserve special attention. In simple model glasses with bare binary interactions, these included quasi-localized modes whose density of states appears to be universal, depending on the frequencies as $D(\omega) \sim \omega^4$, in agreement with the similar law that is obtained with bare forces at zero temperature. In this work, we report investigations of a model of silica glass at finite temperature; here the bare forces include binary and ternary interactions. Nevertheless we can establish the validity of the universal law of the density of  quasi-localized modes also in this richer and more realistic model glass.
\end{abstract}

\maketitle

\section{Introduction}
Simple models of amorphous solids employ ensembles of particles interacting via
binary forces \cite{93KA}. Choosing different sizes of particles (or equivalently, ranges of interaction of these forces), one can create useful models of glass forming systems. In athermal conditions (T=0), these given forces offer also a straightforward path to analyzing the vibrational modes
around a local energy minimum state \cite{99ML}. The bare Hamiltonian $U(\B r_1, \cdots \B r_N)$  provides the Hessian (or force-constant) matrix $\B H$ which determines, in the
harmonic approximation, all the modes and their frequencies \cite{06ML}
\begin{equation}
	H_{ij}^{\alpha \beta} \equiv \frac{1}{\sqrt{m_i m_j}} \frac{\partial^2 U(\B r_1, \cdots \B r_N) }{\partial r_i^\alpha \partial r_j^\beta} \ .
	\label{Hessian}
\end{equation}
Here $\B r_i$ is the $i$th coordinate of a constituent atom of mass $m_i$ in a system with $N$ atoms. As long as the $T=0$ configuration is stable, all the eigenvalues of the
bare Hessian are real and positive (with the exception of few possible zeros associated with Goldstone modes). The force on each atoms $\B F_i$ is given by $-\partial{U(\B r_1, \cdots \B r_N)/\partial \B r_i}$ and it vanishes for all $i$'s in athermal equilibrium.  One then computes the eigenfunctions and eigenvalues of the Hessian $\B H$. The eigenvalues $\lambda_i$ are related to the frequency $\omega_i$ according to
\begin{equation}
	\omega_i =\pm\sqrt{\lambda_i}.
\end{equation}
In amorphous solids, the eigenfunctions can be extended or quasi-localized, with possible hybridization between these classes. In principle, one can distinguish between these different types of modes by considering the participation ratio
$PR$ which is defined as in previous papers \cite{bonfanti2020universal}
\begin{equation}
	PR= [N\sum_i(\B e_i\cdot \B e_i)^2]^{-1} \ ,
	\label{defPR}
\end{equation}
where $\B e_i$ is the $i$th element of a given eigenfunction of the Hessian matrix.
We expect the participation ratio to be of order $O(1/N)$ for a quasi-localized mode (QLM) and of order unity for an extended mode. It was expected for a long time~\cite{91BGGS,03GC,03GPS,07PSG} that the QLM's
display a density of states $D(\omega)$ with a universal power law
\begin{equation}
	D(\omega)\sim \omega^4 \quad \text{in all dimensions.}
	\label{dof}
\end{equation}
However, the actual verification of this prediction was slow in coming. The difficulty is that in large systems the QLM's hybridize strongly with low frequency delocalized elastic extended modes. The latter are expected to follow the Debye theory, with density of states depending on frequency as $\omega^{d-1}$ where $d$ is the spatial dimension. Recently, a remedy was found: By examining {\em small} systems one can bound the frequency of Debye modes from below, exposing the low-frequency QLM's to shine in isolation~\cite{16LDB}. Indeed, in such circumstances
the universal law Eq.~(\ref{dof}) can easily be demonstrated.
A direct verification of such a law with numerical simulations of glass formers with binary interactions~\cite{15BMPP,angelani2018probing,shimada2018spatial,17MSI,18KBL,Moriel2019} and for silica glass with binary and ternary interactions~\cite{bonfanti2020universal,20GRKPVBL} was recently achieved.

Once we turn to finite temperatures, however, it is not immediately obvious how to examine the existence of a similar universal law. The system is never at rest, with atoms moving, colliding, and imparting momentum. The bare Hessian matrix Eq.~(\ref{Hessian}) loses its usefulness, since it generically gains negative eigenvalues when computed
in a given frozen configuration. The total force $\B F_i$ on an $i$th atom, as computed from the bare Hamiltonian,
does not vanish, and the eigenfunctions of the bare Hessian lose their meaning as
modes associated with a frequency of vibration around a well defined
energy minimum. We thus need a new definition of modes that mimics their athermal counterparts.
\begin{figure}
	\centering
	\includegraphics[width=\columnwidth]{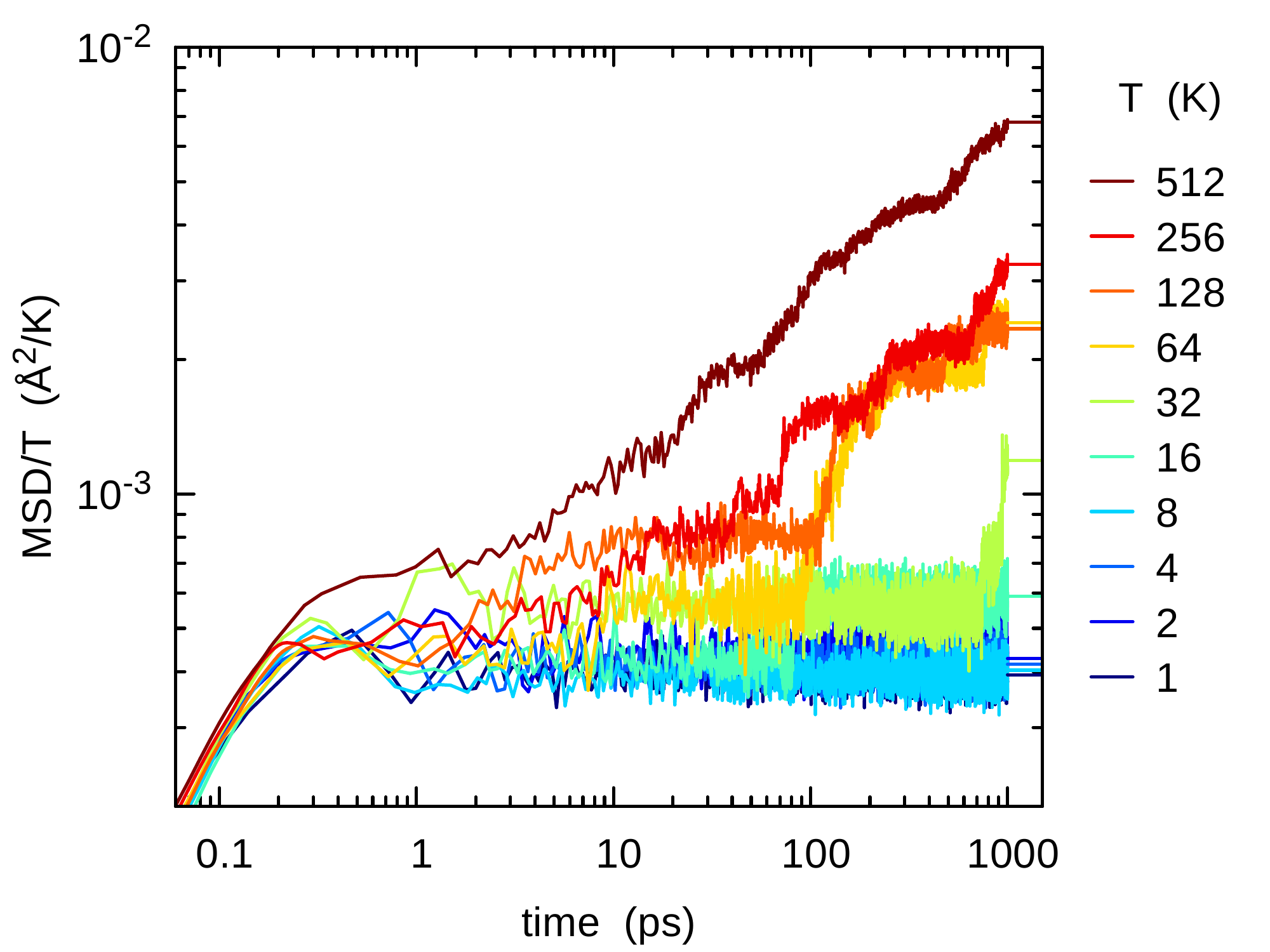}
	\caption{Starting from the inherent structure configuration, we perform MD monitoring the mean squared displacement. Here we plot MSD vs.\ time. In the considered time window, at a large-enough $T$ the cages break and the MSD deviates from the plateau. The final horizontal lines were added to report the final average value of each curve.}
	\label{fig.msd}
\end{figure}

A recently proposed idea focuses on the thermal {\em average} positions of our atoms, and the modes of fluctuations around these~\cite{19DIP,das2021universal}. The average positions of a thermal glass are constant on time scales shorter than the typical diffusion time $\tau_G$.
We thus need to consider relatively stable glasses at sufficiently low temperatures such
that the cage structure around every atom remains stable, apart from thermal motion, for times that are sufficiently long to allow the evaluation of the average position of each atom, but sufficiently shorter than the diffusion time at which the cage structure
is destroyed. At these average positions, the bare forces do not vanish, but one
can consider effective forces which are derived from an effective Hamiltonian that
takes into account the dressing of the forces due to the momentum transfer during
collisions. Of course, these forces will no longer be binary, but rather have ternary, quaternary and higher order contributions \cite{16GLPPRR,19PPSZ}. While it is quite hard to determine precisely the effective forces, it is rather straightforward to define the effective Hessian. To this end, we compute the time averaged positions $\B R_i$:
\begin{equation}
	\B R_i \equiv \frac{1}{\tau}\int_0^\tau dt ~\B r_i(t) \ , \label{defRi}
\end{equation}
where $\tau\ll \tau_G$.
By definition, the positions $\B R_i$ are time independent and the configuration $\{\B R_i\}_{i=1}^N$ is stable, at least within the time interval $[0, \tau_G]$. An additional quantity of importance is the covariance matrix $\B \Sigma$, defined as
\begin{equation}
	\B \Sigma_{ij}\equiv \frac{\sqrt{m_i m_j}}{\tau}\int_0^\tau dt \left(\B r_i(t) - \B R_i\right)\left(\B r_j(t) - \B R_j\right) \ .
	\label{covariance}
\end{equation}

We can now define an effective Hessian via
\begin{equation}
	{\bf H}^{(\rm eff)}=k_B T {\bf \Sigma}^+ \ .
	\label{PDhPI}
\end{equation}
Here ${\bf \Sigma}^+$ is the pseudo-inverse of the covariance matrix \cite{19DIP}.
Next, we note that the effective Hessian given by Eq.~(\ref{PDhPI})  and the covariance matrix have the same set of eigenfunctions
\begin{equation}
	{\bf H}^{(\rm eff)}{\bf \Psi}_i=\lambda^H_i {\bf \Psi}_i
	\label{effeig}
\end{equation}
and their eigenvalues are related by
\begin{equation}
	\lambda^H_i=\frac{k_B T}{\lambda_i^{\Sigma}} \ .
	\label{EigV}
\end{equation}
For all the calculations we have computed the eigenvalues $\lambda^H_i$ from the covariance matrix by taking the inverse of its eigenvalues $\lambda_i^{\Sigma}$ using Eq.~\ref{EigV}, after removing the Goldstone modes.
In Ref.~\cite{19DIP}, it was shown that the eigenvalues and eigenfunctions of ${\bf H}^{(\rm eff)}$ serve the same role for the time-averaged
configuration as the corresponding ones for the bare Hessian play for the athermal configuration. Indeed, in simple model glass formers one could show that the QLM's of
${\bf H}^{(\rm eff)}$ have a universal density of states of the form of Eq.~(\ref{dof}).
The aim of this paper is to examine how universal this result is by studying in
silica glass at non vanishing temperatures.

\section {The model silica glass}
The silica glass is simulated in a 3-dimensional cubic box for two different system sizes:
\begin{itemize}
\item $N$= 1032 atoms, therefore $N_{Si}$= 344 silicon atoms and $N_O$= 688 oxygen atoms, with a box length $L$= 25~\AA
\item $N$= 4008 atoms, therefore  $N_{Si}$= 1336 silicon atoms and $N_O$= 2672 oxygen atoms with a box length $L$= 39.3~\AA,\end{itemize}
The interaction between atoms is given by the Vashishta's potential~\cite{broughton1997direct}.  In this paper, units are defined on the basis of energy, length, and time, respectively being eV, \AA, and ps.

\textit{Preparation Protocol.} Following Ref.~\cite{bonfanti2020universal}, glass samples are initially prepared with randomly positioned Si,O atoms with a density ${\rho}_{in}=2.196$\,g/cm$^3$ and an annealing protocol: (i)~2\,ps of Newtonian dynamics where atoms have Lennard-Jones interactions and are viscously damped with a rate of 1/ps and atomic velocities limited to 1\,\AA/ps, (ii)~8\,ps of damped Newtonian dynamics with Vashishta's potential for silica glass. (iii) Heating up the system up to 4000\,K and then quench to 0\,K in 100\,ps, corresponding to a cooling rate of 40 K/ps. The so-produced configurations are then minimized through the fast inertial relaxation engine (FIRE) \cite{bitzek2006structural} until the total force on every atom satisfies $|\B F_i|\le 10^{-10}$\,eV/{\AA}.\\
\textit{Simulations at non-vanishing temperatures.} We perform simulations using a Langevin thermostat (damping parameter 1\,ps) at $T=1,2,4,8$~K for 50\,ps followed by NVE ensemble simulations for 100\,ps (200\,ps) for the smallest (largest) system size, monitoring the mean square displacements (MSD) of the atoms. The total number of starting configurations for each temperature is 1000.
Differently from our previous works on silica glasses~\cite{bonfanti2018,bonfanti2019}
where we used a different interatomic potential, we use here the Vashishta's potential as implemented in LAMMPS~\cite{lammps} since it is more efficient in terms of computation time.

\section{Results}
To guarantee that our measurements do not exceed the time window in which
diffusion does not play a role, we measure the Mean-Square-Displacement (MSD) of our atoms at each temperature $T$. Since the covariance matrix Eq.~(\ref{covariance}) has to be measured inside the glass basin, we must insure that the system is still in the basin prepared at $t=0$. Figure~\ref{fig.msd} presents the MSD as a function of time.
Obviously, when the temperature is too high, the system escapes from the basin, preventing us from measuring a stationary covariance matrix.
On the other hand for low temperatures (from $T=1$~K up to $T=32$~K), the MSD  reaches a plateau which survives throughout  our simulation window (1000\,ps). Note that as the temperature increases, so the plateau increases, as expected from solid mechanics.  At a finite temperature-dependent timescale $\tau_G$ the MSD departs from the plateau,  meaning that diffusion sets in and the system departs from the local minimum. In practice, we have to compute the covariance matrix within the range of the plateau, before the MSD displays the upturn.

\begin{figure}
	\centering
	~~~~~~~~~~$N=1032$~~~~~~~~~~~~~~~~~~~~~~~~~~~$N=4008$\\
	\includegraphics[width=\columnwidth]{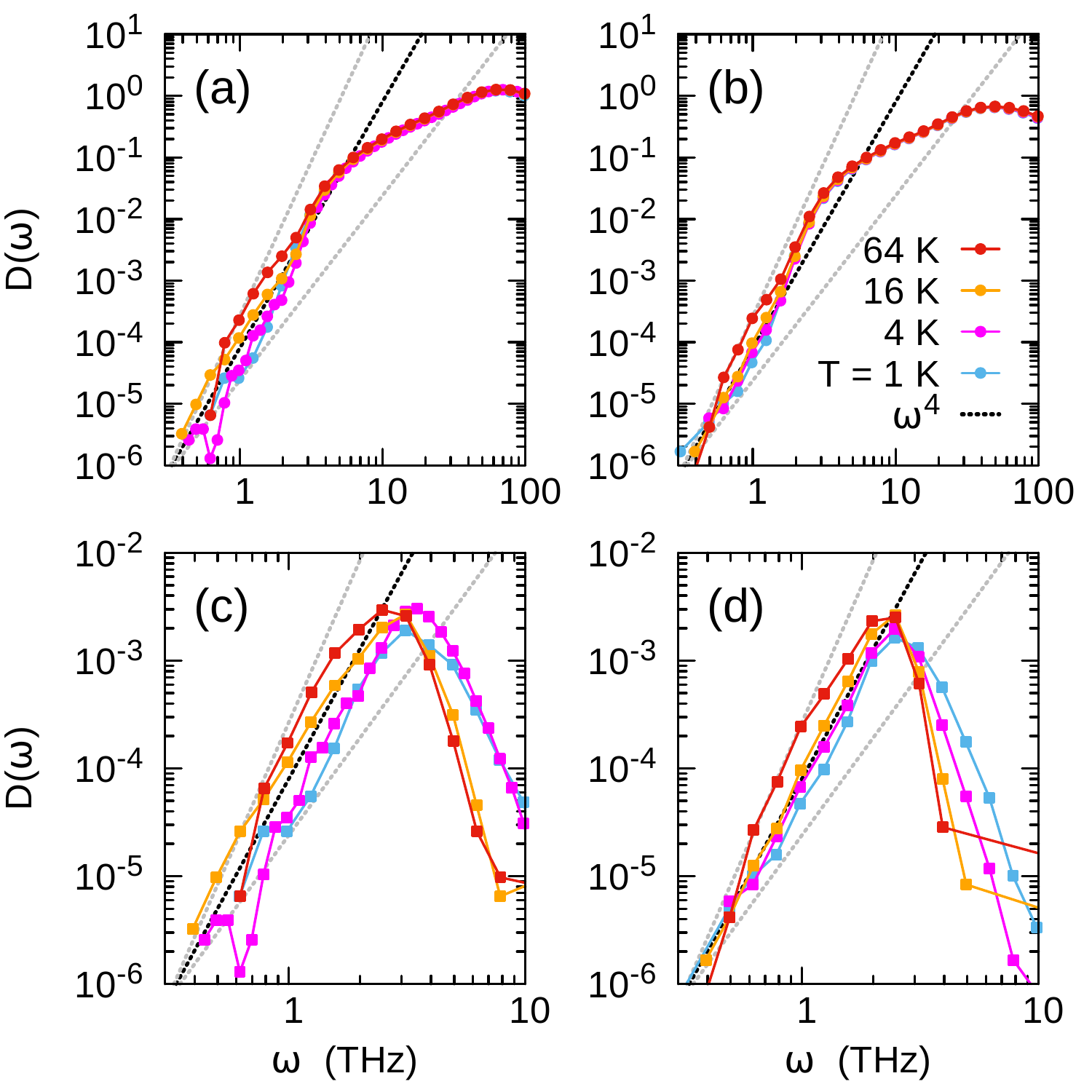}
	\caption{Density of states as computed from the effective Hessian at different temperatures, $T=1,4,16,64$\,K, for SiO$_2$ glass samples of 1032 (panels (a) and (c)) and 4008 (panels (b) and (d)) atoms. In panels (a) and (b) all the eigenfrequencies are included whereas in  panels (c) and (d) only the modes with participation ratio smaller than $0.1$ are considered. For all $T$ values, data from over 1000 samples was included. The grey dotted line report the $\omega^3$ and $\omega^5$ trends.}
	\label{fig.omega4}
\end{figure}

\begin{figure*}
	\centering
	\hspace*{1.1cm} $N=1032$ \hspace*{7cm} $N=4008$\\
	\includegraphics[width=0.9\columnwidth]{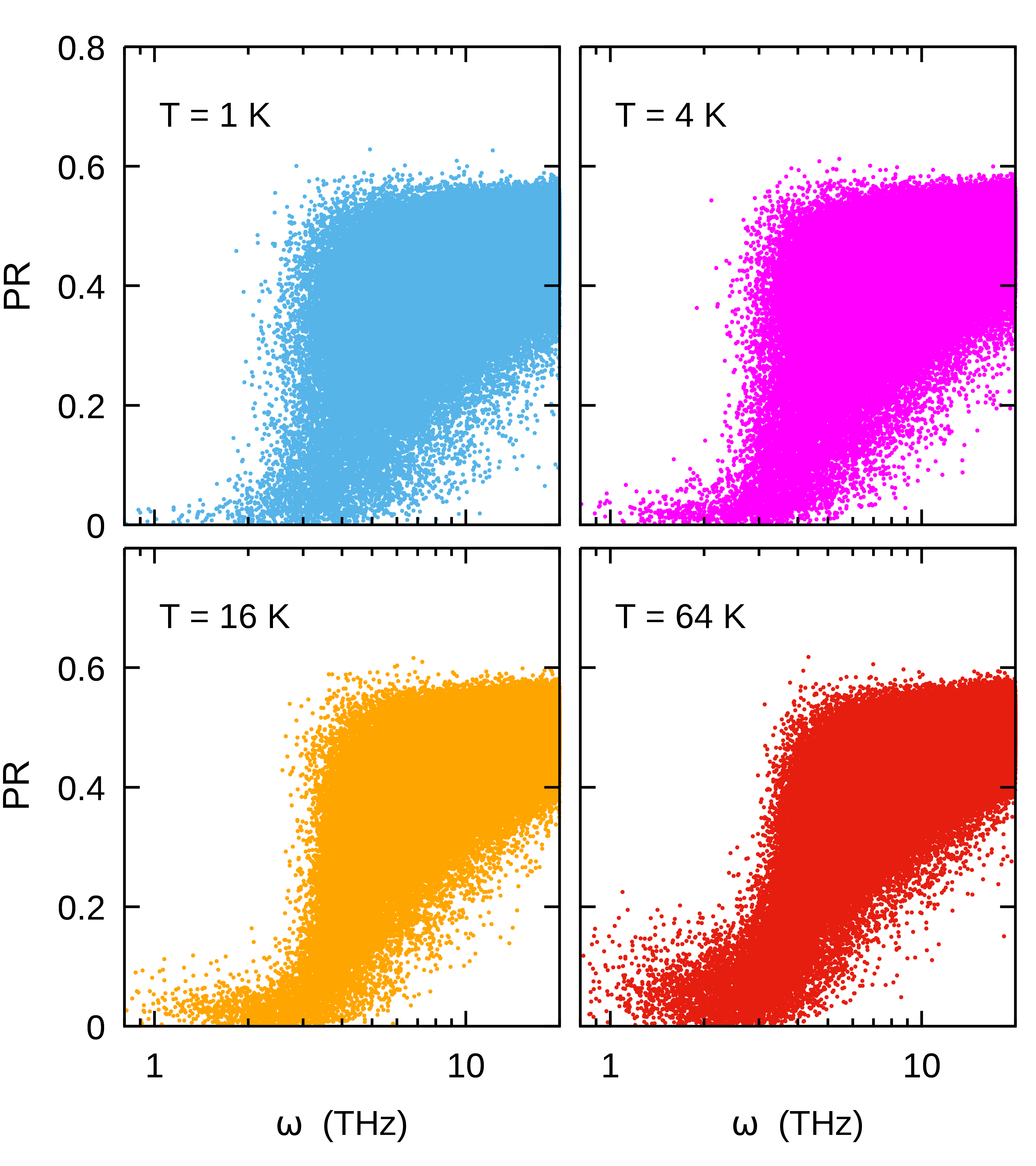}~~~~
	\includegraphics[width=0.9\columnwidth]{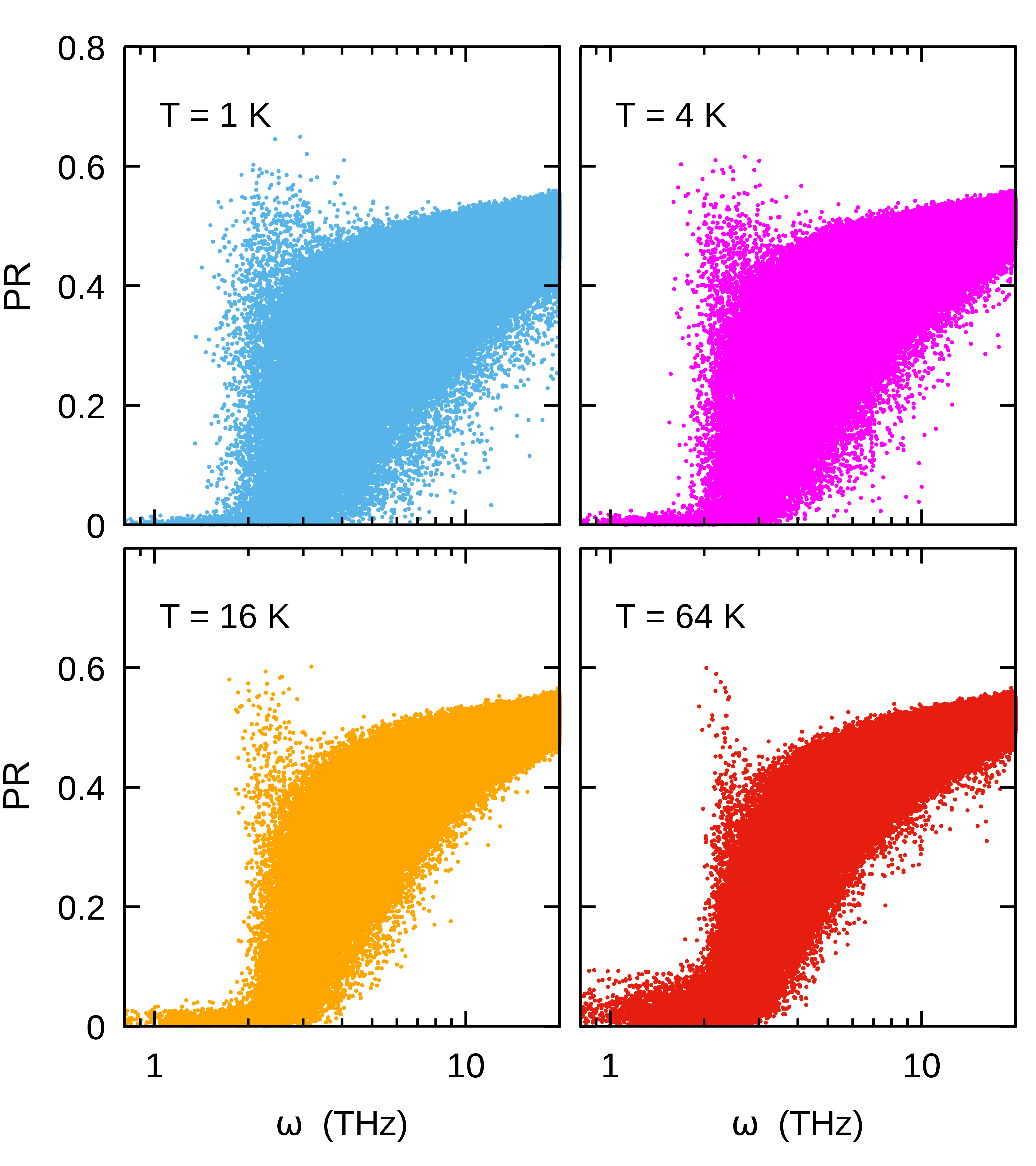}\\
	\caption{
		Participation ratio (Eq.~\ref{defPR}) calculated for the (left panels) $1032$-atoms and (right panels) $4008$-atoms system at four different temperatures $T$.
	}\label{fig.partratio}
\end{figure*}

We show the density of states obtained by the covarince matrix at finite temperatures (from 1\,K to 64\,K) in Figure~\ref{fig.omega4}. Panels (a) and (b) show the density of states including all modes for the two system sizes and the dashed line correspond to the $\omega^4$ scaling law. Only at very low frequency, the DOS appears to obey the $\omega^4$ scaling; for $N=1032$ the behavior is observed in a very short range of frequencies, but for the bigger sample $N=4008$ this trend is clearer. To exhibit the scaling law
more convincingly, we need to select only QLM's. To this end, we include only the modes having participation ratio below 0.1 (see Fig.~\ref{fig.partratio}). Indeed after these modes are selected, we see in panels (c) and (d) that the DOS obeys a clear $\omega^4$ scaling. We note that in the case of  the small system of 1032 atoms (Fig.~\ref{fig.omega4}c) at very low $\omega$, the DOS is much smaller than the expected $\omega^4$ scaling; this is possibly due to finite size effects.
Indeed in the larger samples of 4008 atoms in panel (d), the DOS shows $\omega^4$ scaling down to the very lowest available frequencies.
By fitting $\omega^p$ with the collapsed data of Fig.~\ref{fig.omega4}c,d in the $0.4$--$2.0$\,THz range we have obtained $p = 3.926\pm0.248$ and $p = 3.924\pm0.165$ for $N=1032$ and $N=4008$, respectively.
The data shown in Figure~\ref{fig.omega4} indicates the presence of the $\omega^4$ scaling law also at finite temperatures for the realistic silica glass model. Together with the previously investigated simple binary mixture systems~\cite{das2021universal}, this implies that the $\omega^4$ scaling law is robust and universal, existing in different glass models also at finite temperature.
\begin{figure}
	\centering
	\includegraphics[width=\columnwidth]{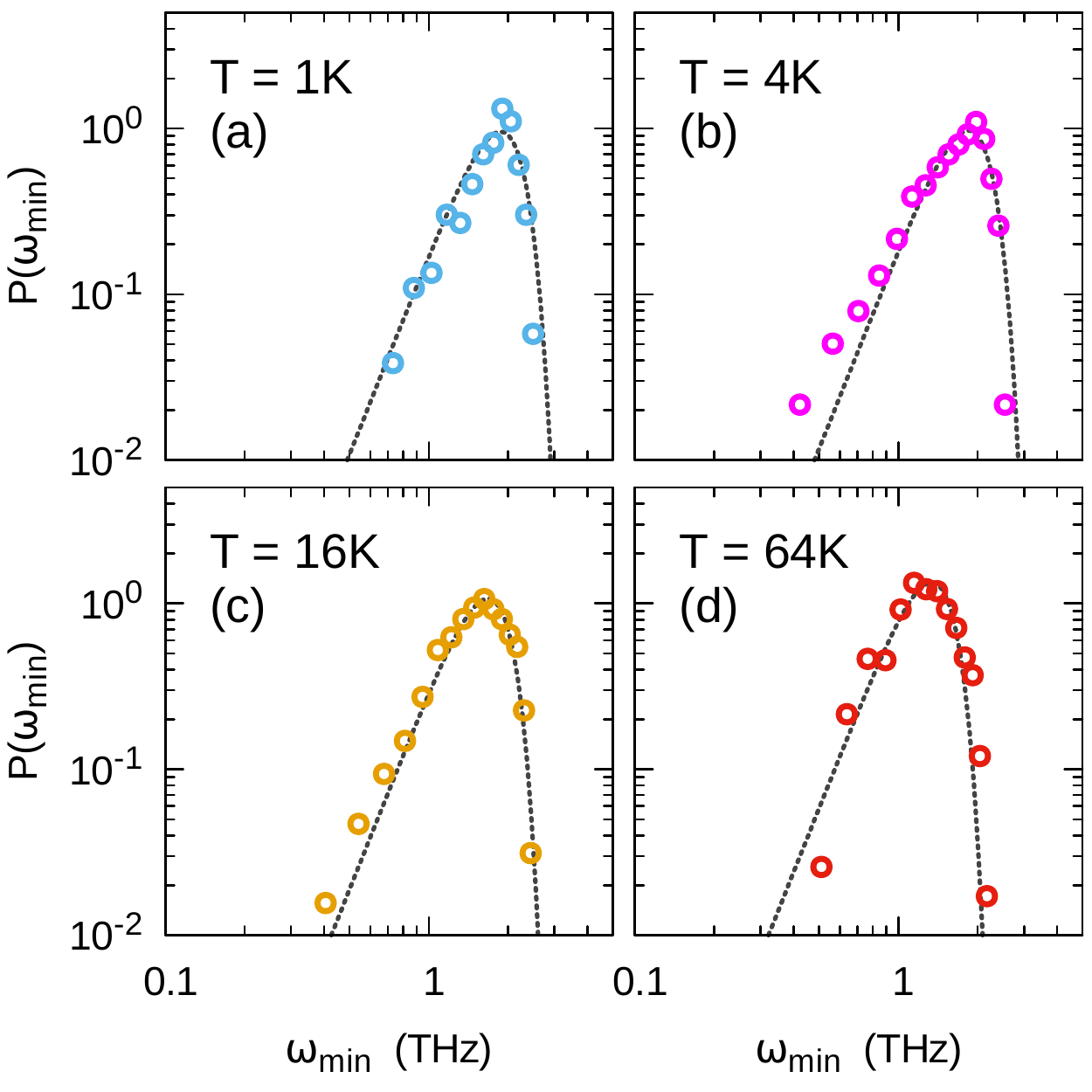}
	\caption{Left panels: Distribution of the minimal vibrational frequency $P(\omega_{min})$ for the largest investigated system size, $N=4008$, and four temperatures $T$. The dotted lines are the corresponding Weibull distributions, Eq.~\ref{Weib}.}
	\label{scaleplot}
\end{figure}

It is interesting to note that recent analysis led two
groups to present density of states with power laws following $\omega^3$ \cite{21WSF} and $\omega^5$ \cite{21KRK} respectively.
Having in mind that in the present case the scaling range of the $\omega^4$ scaling law is rather limited, we turn now to extreme value statistics to lend further support to the $\omega^4$ law.
Since we have many configurations in our simulations, we can determine the minimal frequency obtained from the diagonalization of ${\bf H}^{(\rm eff)}$ in each and every configuration, denoting it as $\omega_{\rm min}$.
The average of this minimal frequency over the ensemble of configurations is denoted $\langle \omega_{\rm min} \rangle$. Referring to the argument first presented in Ref.~\onlinecite{10KLP}, we expect that in systems with $N$ atoms,
\begin{equation}
	\int_0^{\langle \omega_{\rm min}\rangle} D(\omega) d\omega \sim N^{-1} \ .
\end{equation}
Using Eq.~(\ref{dof}), we then expect that in three dimensions
\begin{equation}
	\langle \omega_{\rm min} \rangle \sim N^{-1/5} \sim L^{-3/5} \ .
	\label{mean}
\end{equation}
Moreover, since the different realization are uncorrelated, the values of $\omega_{\rm min}$ are also uncorrelated. Then the well-known Weibull theorem \cite{39Wei} predicts that the distribution of $\omega_{\rm min}$ should obey the Weibull distribution in the limit of
large $N$
\begin{equation}
	W(\omega_{\rm min})=\frac{5(\Gamma(1.2))^5}{\langle \omega_{\rm min}\rangle^5}~\omega_{\rm min}^{4}~e^{-\left(\frac{\omega_{\rm min}\Gamma(1.2)}{\langle \omega_{\rm min}\rangle}\right)^5} \ ,
	\label{Weib}
\end{equation}
where $\Gamma(x)$ is the Gamma function, $\Gamma(1.2)\approx 0.918$.
This prediction is tested in four left panels of Fig.~\ref{scaleplot}.  The distributions of $\omega_{\rm min}$ for four values of the temperature $T$ are shown, together with the predicted distributions as dictated by Eq.~(\ref{Weib}). We stress that there is no free fitting here, and therefore  this is a strong independent test of Eq.~(\ref{dof}). Notice that the Weibull distribution is expected to apply only in the limit of large $N$. Indeed we found that our data for the smaller system with $N=1032$ deviate from the predictions of Eq. (\ref{Weib}).
\begin{figure}[tb!]
\centering
\includegraphics[width=\columnwidth]{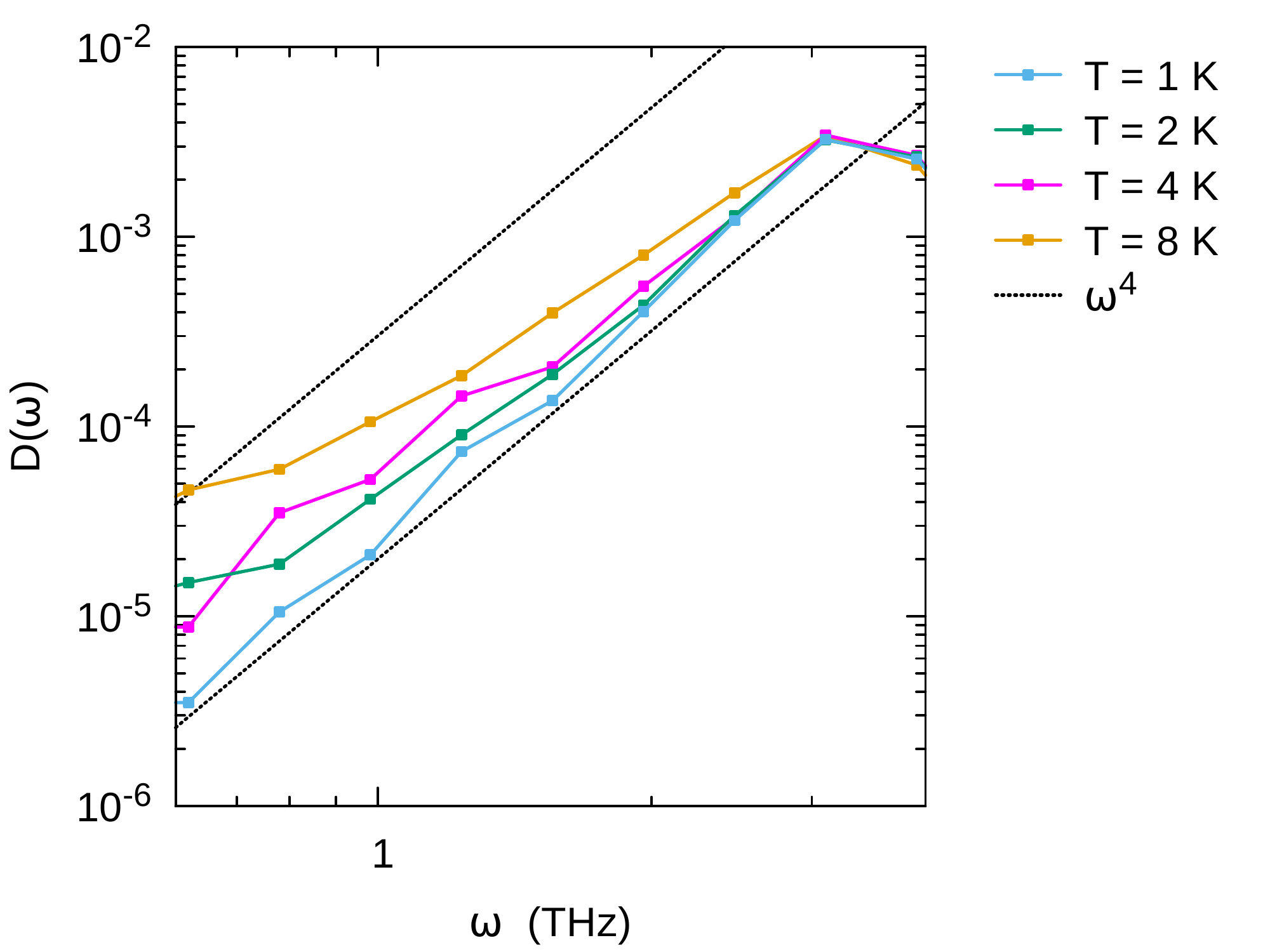}
\caption{Density of states as computed from the standard bare Hessian computed at the mean coordinates $\B R_i$, at different temperatures, $T=1,2,4,8$\,K, for SiO$_2$ glass samples of $N=1032$ atoms. Only the modes with participation ratio smaller than $0.1$ are considered. For all $T$ value, statistics over 1000 samples was accounted.
}
\label{bare}
\end{figure}

\section{Summary and discussion} The main aim of this work was to examine whether the universality class expressed by Eq.~(\ref{dof}) extends to finite temperatures in glasses whose interactions are richer than those of simple glass formers with binary interactions~\cite{das2021universal}. One needs to understand that the bare Hessian, which can be computed for any snapshot of our thermal system, does not yield a scaling law of this form. This bare Hessian has in principle negative eigenvalues (i.e.\ imaginary frequencies), since any given state is unstable and is bound to evolve. In the case of
our silica glass model, when computing the bare Hessian at the mean coordinates $\B R_i$ of Eq.~\ref{defRi}, very low temperature configurations do not have negative eigenvalues. But at higher temperatures the number of negative eigenvalues tends to increase rapidly with $T$; already at $T=16$\,K all the samples showed negative eigenvalues. Therefore, to  consider the density of states one needs to exclude configurations with negative eigenvalues. Selecting these configurations only, and filtering according to the same criterion, i.e.\ including only modes whose participation ratio is smaller than $0.1$, we obtain the density of states shown in Fig.~\ref{bare}. The distribution resembles the unfiltered probability density functions in panels (a) and (b) of Fig.~\ref{fig.omega4}.
To obtain probability density functions following  the scaling law (\ref{dof}), we need to compute ${\bf H}^{(\rm eff)}$ and filter out the modes with a high participation ratio.

{\bf Acknowledgments --} R.G.\ acknowledges financial support from Universit\`a degli Studi di Milano, grant no.\,1094 SEED 2020 - TEQUAD. The work of I.P.\ was supported in part by the US-Israel Binational Science Foundation and the Minerva Foundation, Munich, Germany.

\bibliographystyle{unsrt}
\bibliography{biblio}

\end{document}